\documentstyle[12pt]{article}

\input epsf

\begin{document}

\begin{titlepage}

\title{ Magnetization plateaus in the
ferromagnetic-ferromagnetic-antiferromagnetic Ising chain.}

\author{ V. R. Ohanyan$^{1,2}$\thanks{\it{E-mail address}: vohanyan@www.physdep.r.am },
 N. S. Ananikian$^{1,3}$ \\[1mm]
$^1${\small \sl Department of Theoretical Physics, Yerevan Physics Institute,} \\
{\small \sl Alikhanian Brothers 2, 375036 Yerevan, Armenia}\\[1mm]
$^2${\small \sl Chair of Theoretical Physics, Yerevan State University,} \\
{\small \sl Alex Manoogian st., 1, 375049 Yerevan, Armenia} \\[1mm]
$^3$ {\small \sl   Dipartimento di Scienze Chimiche, Fisiche e
Matematiche,} \\[1mm]
{\small \sl Universita' degli Studi dell'Insubria, via Valleggio,
11-22100 Como, Italy} \\[1mm]}

\maketitle

\begin{abstract}

The Ising chain consisting of the antiferromagneticaly coupled
ferromagnetic trimer is considered in the external magnetic field.
In the framework of the transfer-matrix formalism the
thermodynamics of the system is described. The magnetization per
site $(m)$ is obtained as the explicit function of the external
magnetic field $(H)$. The corresponding plots of $m(H)$ are drawn.
Two qualitatively different regions of the values of coupling
constants are established: weak antiferromagnetic coupling
$\left(J_A<3J_F\right)$ and the strong antiferromagnetic coupling
$\left(J_A\geq3J_F\right)$. For the latter case the magnetization
curve with plateau at $m/m_{sat}=1/3$ is obtained. It is proven
that the plateau is caused by the stability of spatially modulated
spin structure $\langle3111\rangle$. The values of magnetic field
determining the width of the plateau are obtained in the limit of
zero temperature.

\end{abstract}

\end{titlepage}

\section{Introduction}

   Quite recently, the family of known non trivial quantum effects
in the condensed matter physics is enriched with the novel
phenomenon---intermediate plateaus in the magnetization processes.
Namely, the system ceases to responde on changes of the external
magnetic field within some range of the latter. As a consequence,
for some intermediate values of $H$ ($H_1<H<H_2$) a horizontal
region forms in the magnetization ($m$) versus the external
magnetic field ($H$) curve .

 These plateau structures were experimentally observed in series
 of compounds: the triangular antiferromagnets ($C_6Eu$
 \cite{c6eu}, $CsCuCl_3$ \cite{cscucl3}, $RbFe(MoO_4)_2$
 \cite{Rb}) ; quasi one-dimensional $Ni$--compounds\cite{Ni}; quasi
 two-dimensional compound with structure of Shastry-Sutherland
 lattice ($SrCu_2(BO_3)_2$ \cite{Sr}). At present there are
 numerous theoretical papers where the magnetization plateaus
 were obtained for various non-homogeneous Heisenberg chains
 \cite{Hida}-\cite{pl4}, for spin-ladders \cite{ladder},
 and as well as for some two dimensional
 models \cite{2d}-\cite{Hon}.

 In the majority of plateau mechanisms which have been proposed up
 to now the purely quantum phenomena play a crucial role.
 The concepts of magnetic quasiparticles and
 the strong quantum fluctuations are regarded to be of first
 importance for understanding of these processes. Particularly, for a
number of systems it was shown that the plateaus at $m\neq0$ are
caused by the presence of the spin gap in the spectrum of magnetic
excitations in the external magnetic field. Another mechanism lies
in so called crystallization of the magnetic particles \cite{pl1},
\cite{KM}.

Nevertheless, we suppose that the quasiclassical picture in which
the main dynamics of spins can be presented in terms of spin "up"
and spin "down" is also relevant to the problem of magnetization
plateaus. The problem formulated in the quasiclassical language
allows to use the Ising type variable instead of the quantum spin
operators. The advantages of this approximation are obvious: in
the case of one dimensional systems one can develop purely
analytic and quite simple technique for the magnetization plateau
problem.

 One of the pioneering theoretical works where the appearance of the
 magnetization plateau was  predicted is the paper of Hida \cite{Hida}.
 In this paper the
 ferromagnetic--ferromagnetic--anfiferromagnetic Heisenberg chain was
 considered as a model of $3CuCl_2\cdot2$dioxane compound which consists
 of the antiferromagneticaly coupled ferromagnetic trimers. The
 Hamiltonian, considered by Hida has the following form:

 \begin{eqnarray}
 {\mathcal{H}}={\sum_i} {({{{ \mathcal{H}}_i}}^{trim}
 +{{{ \mathcal{H}}_i}}^{int})}+{\mathcal{H}}_Z, \label{1}
 \end{eqnarray}

 where

\begin{eqnarray}
 {{{\mathcal{H}}_i}^{trim}}&=&-2J_F(\mbox{\boldmath{$s$}}_i
 \cdot\mbox{\boldmath{$\tau$}}_i+\mbox{\boldmath{$\tau$}}_i\cdot \label{2}
 \mbox{\boldmath{$\sigma$}}_i),\nonumber \\
{{{\mathcal{H}}_i}^{int}}&=&2J_A(\mbox{\boldmath{$\sigma$}}_i\cdot
\mbox{\boldmath{$s$}}_{i+1}), \label{3}\\
{{\mathcal{H}}_Z}&=&-g{\mu}_BH \sum_{i}({s_i}^z+{{{\sigma}}_i}^z+
\nonumber {{{\tau}}_i}^z), \label{Zeeman}
 \end{eqnarray}

here $J_A$ and $J_F$ are the antiferomagnetic and ferromagnetic
coupling constants respectively, $\mbox{\boldmath $s$}_i$,
 $\mbox{\boldmath $\tau$}_i$ and  $\mbox{\boldmath $\sigma$}_i$ are
 the $S=1/2$ spin operators and ${\mathcal{H}}_Z$ is the standard
 Zeeman Hamiltonian. Using the method of the numerical
 diagonalization of finite size system Hida obtained a
 magnetization curve with the plateau at $m/m_{sat}=1/3$, where
 $m_{sat}$ is the saturation magnetization.

  The approach we are offering in the present paper is based on very
  simple idea to use the Ising spins instead of the Heisenberg operators in
  Eq. (\ref{1}) and to develop the transfer-matrix formalism (see for example
  Ref. \cite{baxter} ). It is worth to emphasize that in contrast to the
  majority of existing approaches to the problem of magnetization plateau,
  our technique is entirely based on analytical calculations and allows
  to obtain the magnetization profiles for arbitrary finite
  temperatures and arbitrary values of parameters $J_A$ and $J_F$.
  As a result we obtained the diagrams of magnetization processes for
  emerged
  ferromagnetic-ferromagnetic-antiferromagnetic Ising chain with the same
  magnetization plateau as in Ref. \cite{Hida}.

Presently, there are the numerical results indicating the
existence of magnetization plateau in two dimensional Ising models
\cite{is}. Moreover, in the Ref. \cite{He} the multisite
interaction Ising model on the infinite dimensional recurrent
lattice was considered. There the magnetization curves with
plateaus at $m=0$ and $m/m_{sat}=1/2$ were also obtained there.
However, till now, there are no any evidence of the appearance of
the magnetization plateaus in the one dimensional Ising chains.

\section{The ferromagnetic-ferromagnetic-antiferromagnetic \\ Ising chain.}
  Replacing all Heisenberg operators in Eq. (\ref{1}) by the Ising
 variables $s, t, u$ one can obtain the model, describing by the follow Hamiltonian:

\begin{eqnarray}
-\beta {\mathcal{H}}=\sum_i (J_F(s_i t_i+t_i u_i)-
J_Au_i s_{i+1}+h(s_i+t_i+u_i)). \label{Ham}
\end{eqnarray}
Here the $J_F=\beta{\tilde{J}}_F$ and $J_A=\beta{\tilde{J}}_A$ are
dimensionless ferromagnetic and antiferromagnetic constants
respectively, $h=\beta H$ is dimensionless magnetic field and all
the variables $s, t, u$ take values equal to $\pm1$.
 The partition function for one dimensional systems can be written
 in the following way:
\begin{eqnarray}
Z=\sum_{(s,t,u)} e^{-\beta {\mathcal{H}}}=\sum_{(s,t,u)}\prod_i
\exp(J_F(s_i t_i+t_i u_i)-J_Au_i s_{i+1}+h(s_i+t_i+u_i)).  \label{PF}
\end{eqnarray}
Here the product is going over all sites of the chain. If the
periodic boundary conditions are assumed, the partition function
can be represented as a trace of product of the transfer-matrices:

\begin{eqnarray}
Z=\mathbf{Sp}(\mathbf{T}^N). \label{spur}
\end{eqnarray}
In our case we can write down the elements of the transfer-matrix
as

\begin{eqnarray}
T_{s\acute{s}}=\sum_{t,u} exp(J_F(s t+t u)-J_{AF}u
\acute{s}+h(s+t+u)) =\sum_{t,u}
exp({\alpha}_{1}(t)s+{\alpha}_{2}(u)\acute{s}
 +{\alpha}_{3}(t,u)), \label{tm}
\end{eqnarray}
where the following notations are introduced:
\begin{eqnarray}
\alpha_1=J_Ft+h, \quad \alpha_2=-J_Au,
\quad \alpha_3=J_Fut+h(t+u). \label{alpha}
\end{eqnarray}
 Implementing the summation on $t$ and $u$ in Eq. (\ref{tm})
  one can easily obtain the
 explicit form of the transfer-matrix:
\begin{eqnarray}
\mathbf{T}= \left( \begin{array}{lcr}
      2e^{J_A}\cosh h+2e^{-J_A+2h}\cosh (2J_F+h) &
      2e^{-J_A}\cosh h+2e^{J_A+2h}\cosh (2J_F+h)\\
      2e^{-J_A}\cosh h+2e^{J_A-2h}\cosh (2J_F-h) &
      2e^{J_A}\cosh h+2e^{-J_A-2h}\cosh (2J_F-h)\label{TM}
      \end{array}
\right).
\end{eqnarray}
 Following the general procedure, using the properties of the matrix
 trace, $\mathbf{Sp}(A_1 A_2 \ldots A_kB)
 =Sp(BA_1 A_2\ldots A_k)$ it is easy to show
 that the partition function from Eq. (\ref{PF}) for
 one-dimensional model is the sum of the $N$-th power
 of the eigenvalues of transfer-matrix.
 \begin{eqnarray}
 Z=\sum_i {\lambda_i}^N. \label{pfl}
 \end{eqnarray}
 It is easy to see that in the thermodynamical limit, $N \to \infty$,
 only the maximal eigenvalue
 contributes to the r. h. s. of Eq. (\ref{pfl}). Thus, the problem of finding the partition
 function for the chain reduces to the problem of finding the maximal
 eigenvalue of the transfer-matrix. Solving the secular equation for the
 matrix (\ref{TM}), one can obtain the eigenvalues
\begin{eqnarray}
&\lambda_{\pm}&= 2\cosh h+e^{-J_A}\Big(
e^{2h}\cosh(2J_F+h)+e^{-2h}\cosh(2J_F-h)\pm\\ \nonumber &\pm&
\Big((e^{2h}\cosh(2J_F+h)-e^{-2h}\cosh(2J_F-h))^2 \\ \nonumber
&+&4(\cosh h+ e^{2(J_A+h)}\cosh(2J_F+h))(\cosh
h+e^{2(J_A-h)}\cosh(2J_F-h))\Big)^{1/2}\Big).
\end{eqnarray}
 It is easy to see that for our purpose we have to take $\lambda_+$.

  Thus, in the thermodynamical limit the free energy per site takes the form:

 \begin{eqnarray}
 &F&=\lim_{N\to\infty}\left(-\frac{\log{{\lambda}_+}^N}{3N\beta}\right)= \\ \nonumber
 &=&-\frac{1}{3\beta}
\log\bigg\{2\cosh h+e^{-J_A}\Big(
e^{2h}\cosh(2J_F+h)+e^{-2h}\cosh(2J_F-h)+\\ \nonumber &+&
\Big((e^{2h}\cosh(2J_F+h)-e^{-2h}\cosh(2J_F-h))^2 \\ \nonumber
&+&4\big\{\cosh h+ e^{2(J_A+h)}\cosh(2J_F+h))(\cosh
h+e^{2(J_A-h)}\cosh(2J_F-h)\big\}\Big)^{1/2}\Big)\bigg\} .
\label{F}
 \end{eqnarray}
 The factor $1/3$ is introduced here because there are three spins in each
 site of the chain, and the total number of spins is $3N$. Now, using
 the general thermodynamical relations we can find the magnetization
 per site
 \begin{eqnarray}
 m=-\left(\frac{\partial F}{\partial H}\right)_{\beta}=
 \frac{1}{3\beta {\lambda}_{+}}\left(\frac{{\partial {\lambda}_{+}}}{{\partial H}}\right)
_{\beta}=\frac{1}{3{\lambda}_{+}}\left(\frac{{\partial
{\lambda}_{+}}}{\partial h}\right)_{\beta}. \label{m}
 \end{eqnarray}
 The explicit form of this expression is too cumbersome and we
found no reason to write it down here. However, it is quite
suitable to plot analytically the corresponding magnetization
profiles (diagrams of $m$ versus the external magnetic field $H$)
 with the aid of MATHEMATICA 4.0 package. The results
are presented in the next section.

\section{Magnetization Processes.}

It is easy to see that the ground state of the system under
consideration at $T=0$ and $H=0$ is the antiferromagnetic
spatially modulated structure in which the trimers of spins
pointed up are alternating with the trimers of spins pointed down
($\ldots
\uparrow\uparrow\uparrow\downarrow\downarrow\downarrow\uparrow
\uparrow\uparrow\downarrow\downarrow\downarrow \ldots$ and so on).
Usually this modulated phase is denoted by $\langle3\rangle$
\cite{mod}.
  This is valid for all values of $J_A$ and $J_F$ in
the absence of the external magnetic field \footnote {Hereafter
for convenience we will identify $\tilde{{J_A}}$ by $J_A$ and
$\tilde{{J_f}}$ by $J_F$.}.

 The profile of the magnetization as a function of the external
 magnetic field essentially depends on the ratio of coupling
 constants $\kappa=J_A/J_F$ and temperature. Actually one can
 distinguish two regions of values of $\kappa$: strong antiferromagnetic
 coupling and weak antiferromagnetic coupling. The limiting case
 of the latter, when
 the system reduces to the ordinary ferromagnetic Ising chain is $J_A=0$.
  When the antiferromagnetic coupling $J_A$
is included then the low temperature behaviour changes steeply. At
$T=0$ and $H=0$ as is mentioned above the chain is in the ordered
state with complex antiferromagnetic long--range order. At finite
but sufficiently low temperatures the long--range order is
breaking just by few solitons. The external magnetic field will
have only small effects until it reaches a critical value $H_c$
where it can flip a spin without cost in energy. Thus, the
appearance of a magnetization plateau in the magnetization versus
the external magnetic field curve is expected in this case. The
width of the plateau is obviously equal to $H_{c_1}$. For the
chain at $T=0$ one, using the simple principles of minimal energy,
can easily express $H_{c_1}$ in terms of $J_A$ and $J_F$.

 Let us first consider the case when antifferromagnetic coupling
 between trimers is assumed to be so intensive that under the effect of
 the sufficiently low external magnetic field only the central spins of
 ferromagneticaly coupled trimers flip (do responde);
 whereas the rest two spins are maintained in the same position as
 in the ground state. This regime we will call the strong
 antiferromagnetic coupling. So, in this case, when the magnitude
 of the external magnetic field reaches its critical value
 $H_{c_1}$ the system will pass from its ground state
  $\langle3\rangle$ to the novel spatially modulated structure
 $\langle3111\rangle$, in which the periodic sequence of spins
 consists of one trimer of spins pointed along the field and another
 trimer with alternating orientation of spins (See Fig. 1). In
 order to obtain the value of the critical field $H_{c_1}$ one
 must compare the energies of $\langle3\rangle$ and
 $\langle3111\rangle$ states and find the value of $H$
 at which the  $\langle3111\rangle$ state becomes energetically
 more preferable than the ground state. Considering the single
 periodic spin sequences we have

 \begin{eqnarray}
 -J_A-2H\leq-4J_F-J_A,\quad \textstyle{or}\quad H_{c_1}=2J_F. \label{hc1}
 \end{eqnarray}

 So, for the strong antiferromagnetic coupling regime the value of
 the critical field does not depend on $J_A$.

  Now we can give a quantitative criterion for this domain of
  parameters. The antiferromagnetic interaction between edge spins
 of trimers is strong in the sense described above when the flip of
 the central spin does not lead to the flip of its adjacent one
 at the critical value of the external field calculated above. In
 the other words, the values of coupling $J_A$ and $J_F$ are such
 that the $\langle3111\rangle$ structure is energetically more
 preferable than the spin configuration where at least one of the spins
 adjacent  to the central one is pointed up at $H=H_{c_1}$.
 Comparing corresponding energies for the single periodic sequence
 one can obtain the following condition of strong ferromagnetic
 coupling (Fig. 1):

 \begin{eqnarray}
 -2J_A-2H_{c_1}\leq-2J_F-4H_{c_1}, \quad \textstyle{or} \quad
 J_A\geq3J_F. \label{con}
 \end{eqnarray}

The weak antiferromagnetic coupling, $J_A<3J_F$, supposes that the
ferromagnetic interaction inside the trimers is strong enough, so
that the three spins the trimer consists of behave like single
Ising spin, which can take values $\pm3$ with effective
antiferromagnetic coupling $J_{eff}=J_A/3$. This case corresponds
to the $S=3/2$ singlets in the quantum model considered by Hida
\cite{Hida}. Thus, for $J_A<3J_F$ the system under consideration
can be regarded as the simple antiferromagetic Ising chain with
the restrictions given above. The value of the critical field
magnitude can be easily obtained in this case. Obviously,
$H_{c_1}=2J_{eff}=\frac{2}{3}J_A$, thus,

\begin{eqnarray}
H_{c_1}=\left\{ \begin{array}{rl}
            2J_F, & \mbox{if } J_A\geq3J_F \\
            \frac{2}{3}J_A, & \mbox{if } J_A<3J_F.
             \label{hcc}
\end{array} \right.
\end{eqnarray}

 The plots, presented in the Fig. 2 illustrate these results.
 Note,
 that the high temperature magnetic behaviour for all regions of
 parameters is rather uniform and the typical magnetization
 profile is given by the curve of Langevin type. So,
 in that follows we will consider only low temperature region.
 At finite temperatures the values of $H_{c_1}$ are obviously
 different from Eq. (\ref{hcc}) due to existence of thermal fluctuations
 of spins. This difference becomes smaller with the decrease of
 temperature. So, for sufficiently low temperatures
 \begin{eqnarray}
 H_{c_1}(T)=H_{c_1}-\theta(T),\quad \lim_{T\to0}\theta(T)=0.
 \label{t}
 \end{eqnarray}
 Fig.2(a)-2(c) show that already at $T=0.1J_F$ value of critical
 field is quite close to $\frac{2}{3}J_A$.

  The regime of strong antiferromagnetic coupling is
  characterized by additional feature which is absent in case of
  $\kappa<3$. It is the magnetization plateau at
  $m/m_{sat}=1/3$. Corresponding plots are presented in Fig. 3.
  One can easily see that the width of plateau depends on the
  value of $\kappa$, namely it becomes broader with the increase of
  $\kappa$. In this case another critical value of the
  external magnetic field magnitude defining the width
  of the plateau appears. So, in the magnetization curve one can see a
  horizontal region for the values of $H$ in interval
  $[H_{c_1},H_{c_2}]$. With the aid of calculations
  analogous to those we made to obtain $H_{c_1}$ one can easily
  get
  the second critical value $H_{c_2}$ for the chain at T=0 in the
  strong antiferromagnetic coupling region. Appearance of the
  plateau at $m/m_{sat}=1/3$ and $H=H_{c_1}$ is indication of the
  stability of $\langle3111\rangle$ configuration when $H\in[H_{c_1},H_{c_2}]$
  . So, when $H$ reaches its second critical
  value $H_{c_2}$ the above mentioned structure begins to destroy
  due to further flip of the spins which previously were pointed down. The
  comparison of the corresponding energies of single periodic spin
  sequence yields (Fig. 1):
  \begin{eqnarray}
  -2J_F-4H\leq-2J_A-2H, \quad or \quad H_{c_2}=J_A-J_F, \label{hc2}
  \end{eqnarray}
  this is in full agreement with Fig.3. The effects of finiteness
  of temperature play the same role as in the case of $H_{c_1}$ i.
  e. they can only bring a slight decrease of the $H_{c_2}$ at
  low values of $T$. It is worth to note once more that high temperatures
  always lead to disappearance of plateau due to strong
  intensity of thermal fluctuations of spins which in their turn
  destroy the ordered configuration. From all stated above we can
  conclude that the ferromagnetic-ferromagnetic-antiferromagnetic
  trimerized Ising chain in the region of strong antiferromagnetic
  coupling have a magnetization plateau, its width in the limit
  of zero temperature is equal to $H_{c_2}-H_{c_1}=J_A-3J_F$.

  In the Fig. 4 the plots of magnetization for given values of
   coupling $(\kappa=6)$ and different values of temperature are shown. At $T=2.5J_F$
    the curve
    has no deviations from the usual high temperature
    magnetic behaviour of spin systems (Fig. 4(a)).
But with decrease of temperature some peculiarities appear in the
course of magnetization curve. In Fig. 4(b) one can see the first
signs of forthcoming plateaus at $m=0$ and $m/m_{sat}=1/3$ and the
complete formation of these plateaus at $T=0.1J_F$ (Fig. 4(c)).
Fig. 4(c) also clearly indicates that the values of critical
fields are just slightly different from $H_{c_1}$ and $H_{c_2}$
calculated above. And finally the curve, presented in Fig. 4(d),
obtained for extremely low temperature $T=0.001J_F$ shows
step--like behaviour which is typical for $T=0$. In this case, one
can see that the jumps in the course of the magnetization exactly
coincide with $H=2J_F$ and $H=J_A-J_F$ that is in full agreement
with the pattern of transition from $\langle3\rangle$ state to
$\langle3111\rangle$ state with further transition to the
saturated state. Nevertheless, it is necessary to emphasize that
there are no real jumps at any finite temperatures at all, because
the system under consideration is one--dimensional and it is know
that there are no phase transition in classical systems in $d=1$.
So, the results presented in Fig. 4(d) should be regarded not as
jumps of magnetization but as extremely steep rises of the $m(H)$
function in the vicinity of critical values of $H$.

\section{Conclusion.}

 The main aim being pursued in this paper is to
 establish the formation of magnetization plateaus in the
 classical one-dimensional spin systems with Ising interaction. As
 an example we chose the classical counterpart of the
 ferromagnetic-ferromagnetic-antiferromagnetic trimerized
 Heisenberg chain, considered by Hida \cite{Hida}. The approach
 developed by Hida was based on a numerical diagonalization of
 finite size system and allowed him to explain the low temperature
 magnetization data for the compound $3CuCl_2\cdot2dx$ as well as
 to predict the appearance of a magnetization plateau at
 $m/m_{sat}=1/3$ when the ferromagnetic exchange energy is
 comparable to or smaller than the antiferromagnetic exchange
 energy. Using entirely analytic technique of the transfer matrix
 for corresponding system of Ising spins we have obtained the
 exact explicit expression for magnetization function.
Despite the extremal complexity of the expression, nevertheless it
permits to draw the plots of magnetization processes for arbitrary
finite temperature and couplings.
  Obtained plots completely reproduce the result of
 Hida, moreover in our model the exact condition of
 existence of plateau $J_A\geq3J_F$ is derived. It is also
 possible using our approach to calculate the width of the plateau
 at $m/m_{sat}=1/3$. The latter is equal to $J_A-3J_F$ in the
 limit $T\to0$.

  The phenomenon of magnetization plateaus are often regarded to
  have purely quantum origin, however it is obviously not always
  true since there are some two dimensional classical spin models
  which exhibit a magnetization plateaus at zero temperature. For
  instance, in Ref. \cite{2d} Kubo and Momoi investigated the
  Multiple Spin Exchange (MSE) model on the triangular lattice in
  the classical limit and have found a large range of parameters
  where magnetization plateaus at $m/m_{sat}=1/3$ and
  $m/m_{sat}=1/2$ occur at zero temperature. There are also
  results on the two dimensional Ising models \cite{is} as well as
  on the calculations on recurrent lattice \cite{He}.
 However, for one dimensional
  systems these results are obtained for the first time.

   Moreover,
  these allow to suppose
  that there exists a class of one dimensional spin systems which
  being both considered in terms of Heisenberg and Ising spins
  leads to the qualitatively same structures of the magnetization
  profiles, particularly to the formation of magnetization
  plateaus. In contrast to the quantum systems the plateau in our
  model has very simple nature associated with the stability of
  spatially modulated structure, so called $\langle3111\rangle$
  configuration of spins which we proved to exists at $T=0$.
  Apparently, for finite but sufficiently low temperatures the plateau is
  still caused by the same $\langle3111\rangle$ structure,
  though the other configurations are also possible, for instance so
  called $uud$ phase which is responsible for plateau at
  $m/m_{sat}=1/3$ in triangular antiferromagnets and MSE model.

    In conclusion note, that the approach based on Ising spins
   is remarkable for its simplicity and clarity. It doesn't
   require any numerical calculations in contrast to the majority of
   methods, which are used in this field of research, and can serve as a very
   useful supplement to the generally applicable techniques such as the numerical
   diagonalizations of finite clusters, Quantum Monte--Carlo
   calculations, series expansions and bosonization technique.

\section{Acknowledgments.}
VRO would like to thank for hospitality the Joint Institute of
Nuclear Research in Dubna where the main part of the paper was
done. The authors are grateful to Prof. V. Priezzhev, Vl. Papoyan
and Prof. V. Morozov for fruitful discussions, Prof. M. Roger for
valuable comments and H. Asriyan for careful reading of the paper.
This work was partially supported by Cariplo Foundation.

\newpage
\begin{center}
 \Large
 Figure Capture
 \end{center}
 \vspace{1cm}

 \vspace{1cm}

 \hspace{0.5cm}Fig. 1 Periodic spin sequences and corresponding
 energies for (a): ground state configuration at $T=0$ and $H=0$;
 (b): the plateau state, so called $\langle3111\rangle$
 configuration; (c): one of the possible intermediate
 configurations which forms after the plateau state. \\
  Double line between the
 sites denote the antiferromagnetic interaction $J_A$.
 \vspace{1cm}

 \hspace{0.5cm}Fig.2 Magnetization curves in the regime of weak
 antiferromagnetic coupling, $J_A\leq3J_F$ at $T=0.1J_F$ for different values of $\kappa$:
 (a) $\kappa=1$; (b) $\kappa=2$; (c) $\kappa=3$.\\
  The value of the
 critical field magnitude in these cases is $3/2J_A$.

 \vspace{1cm}

 \hspace{0.5cm}Fig.3 Magnetization curves in the regime of strong
 antiferromagnetic coupling exhibiting the magnetization plateau
 at $m/m_{sat}=1/3$ for $\kappa=4$ (a), $\kappa=5$ (b) and
 $\kappa=10$ at $T=0.1J_F$. The values of critical field
 magnitude, determining the width of plateau are $H_{c_1}=2J_F$
 and $H_{c_2}=J_A-J_F$.

 \vspace{1cm}

 \hspace{0.5cm}Fig. 4 The magnetization curves for $\kappa=6$
 at different temperatures: (a) $T=2.5J_F$; (b) $T=0.8J_F$; (c)
 $T=0.1J_F$; (d) $T=0.001J_F$.

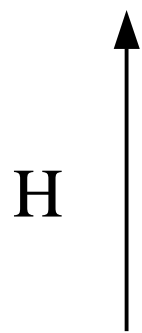
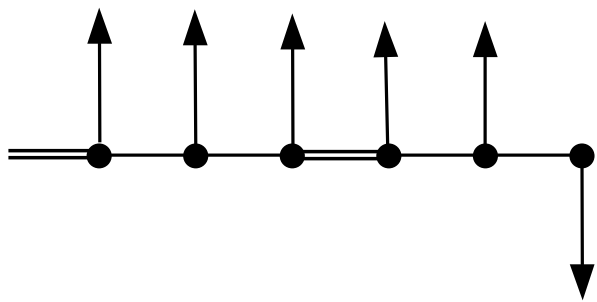
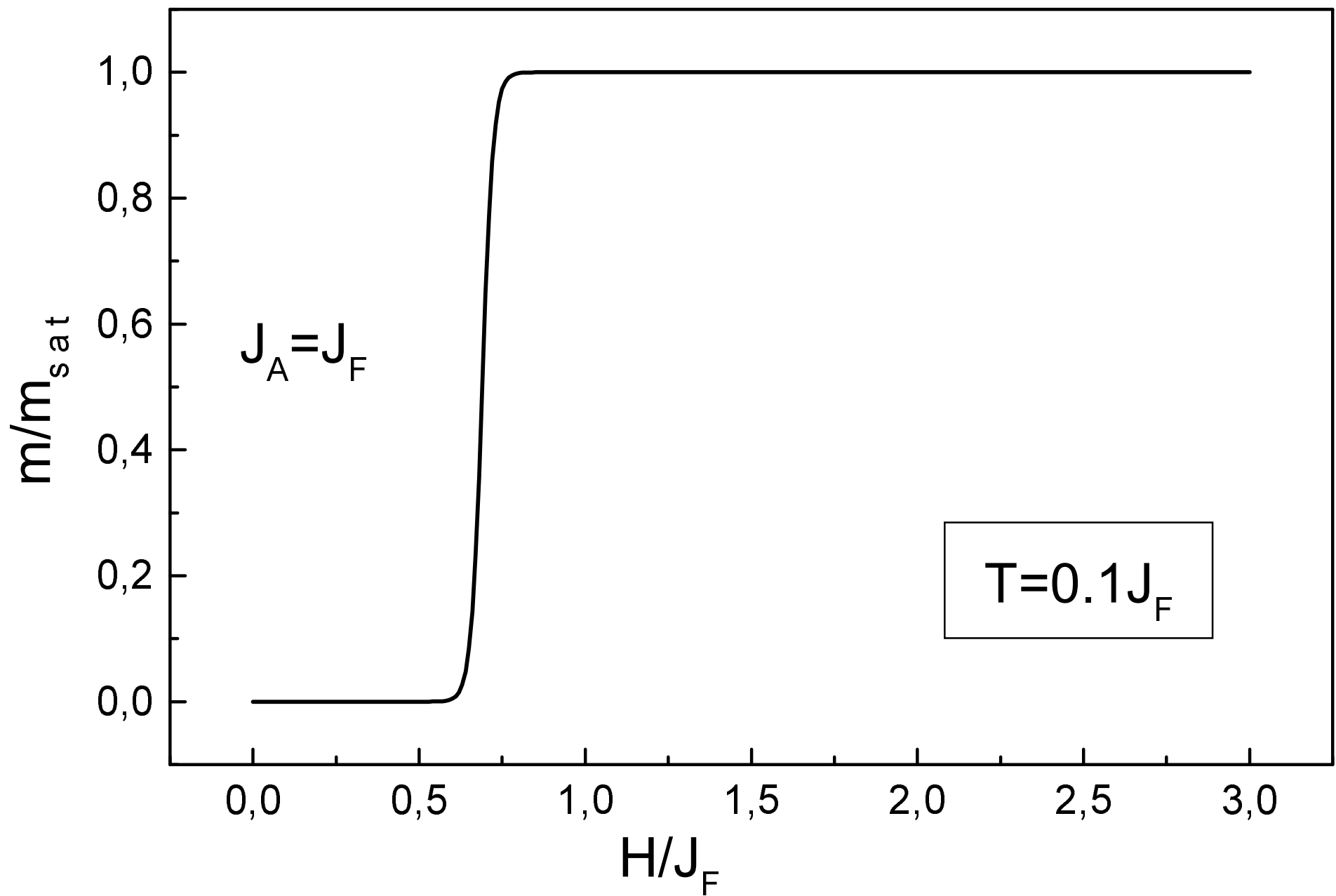
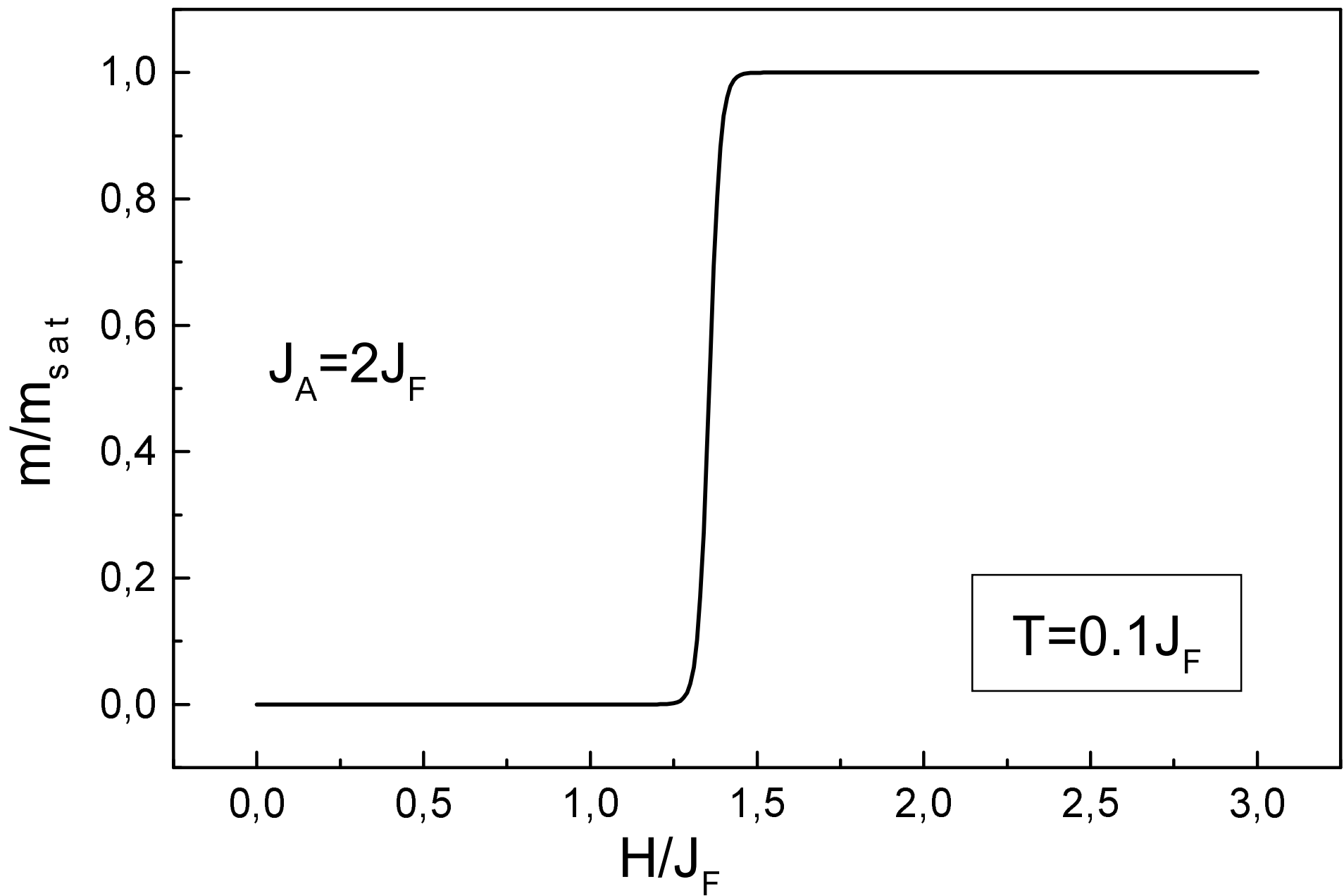
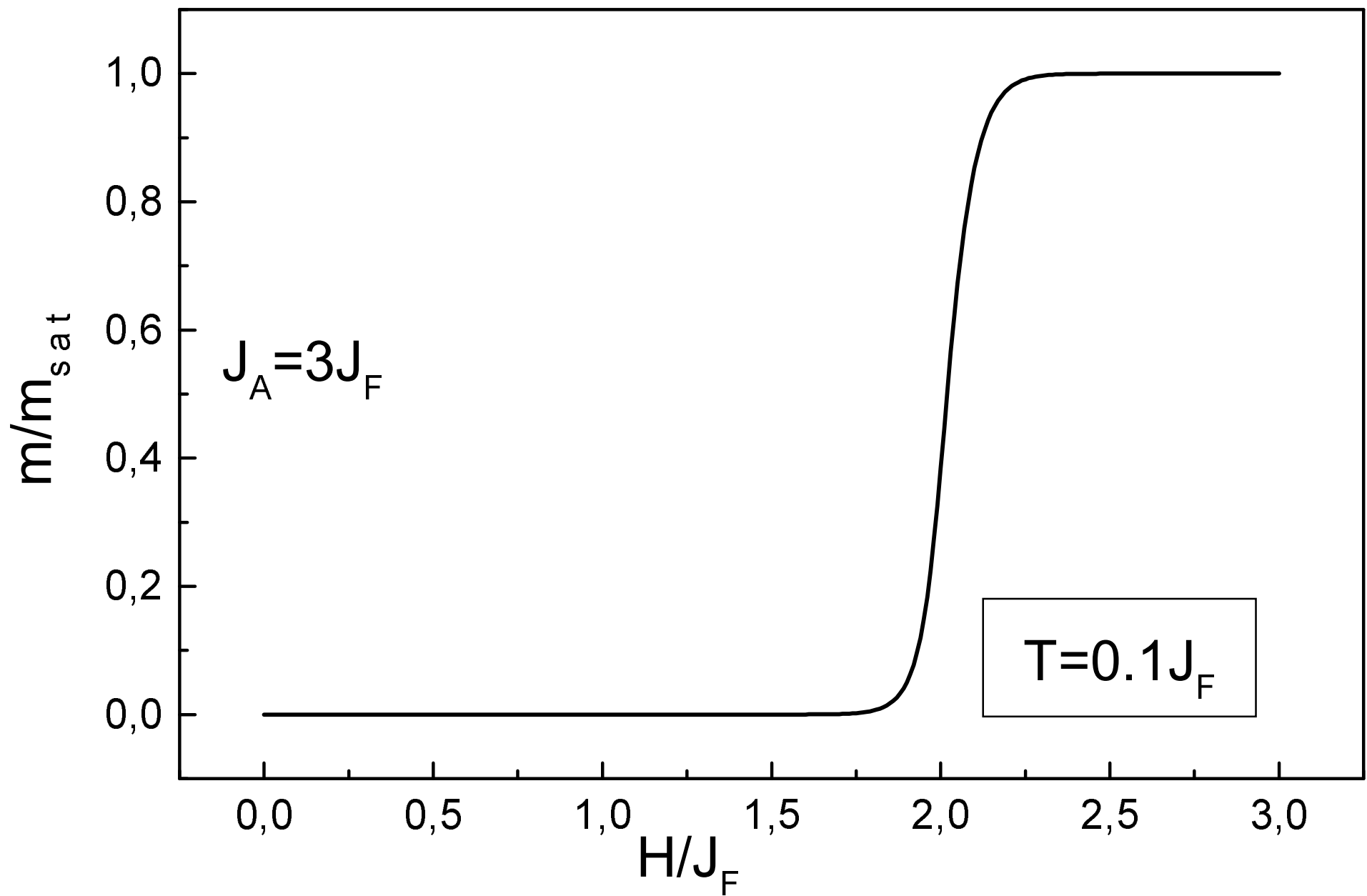
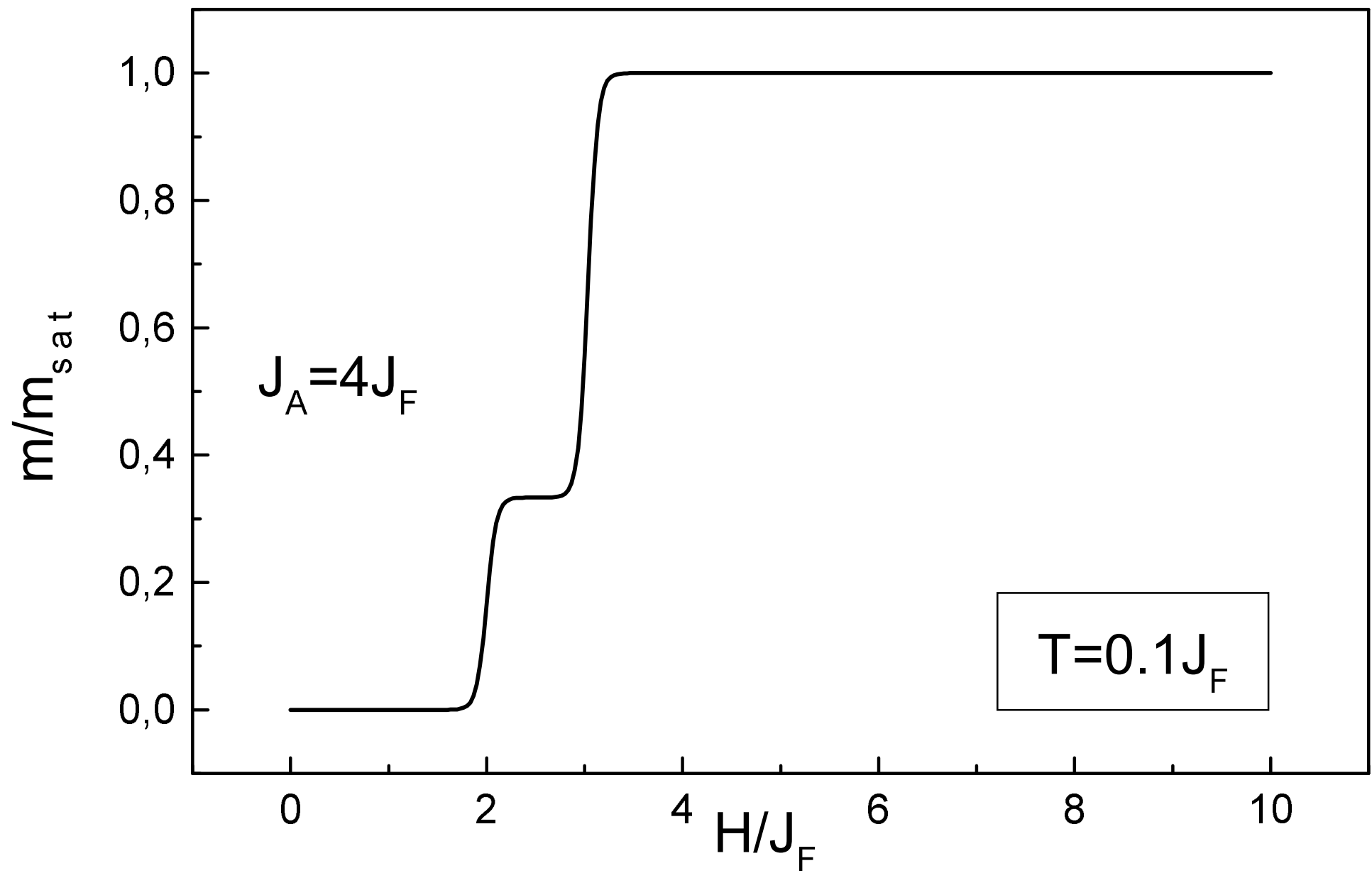
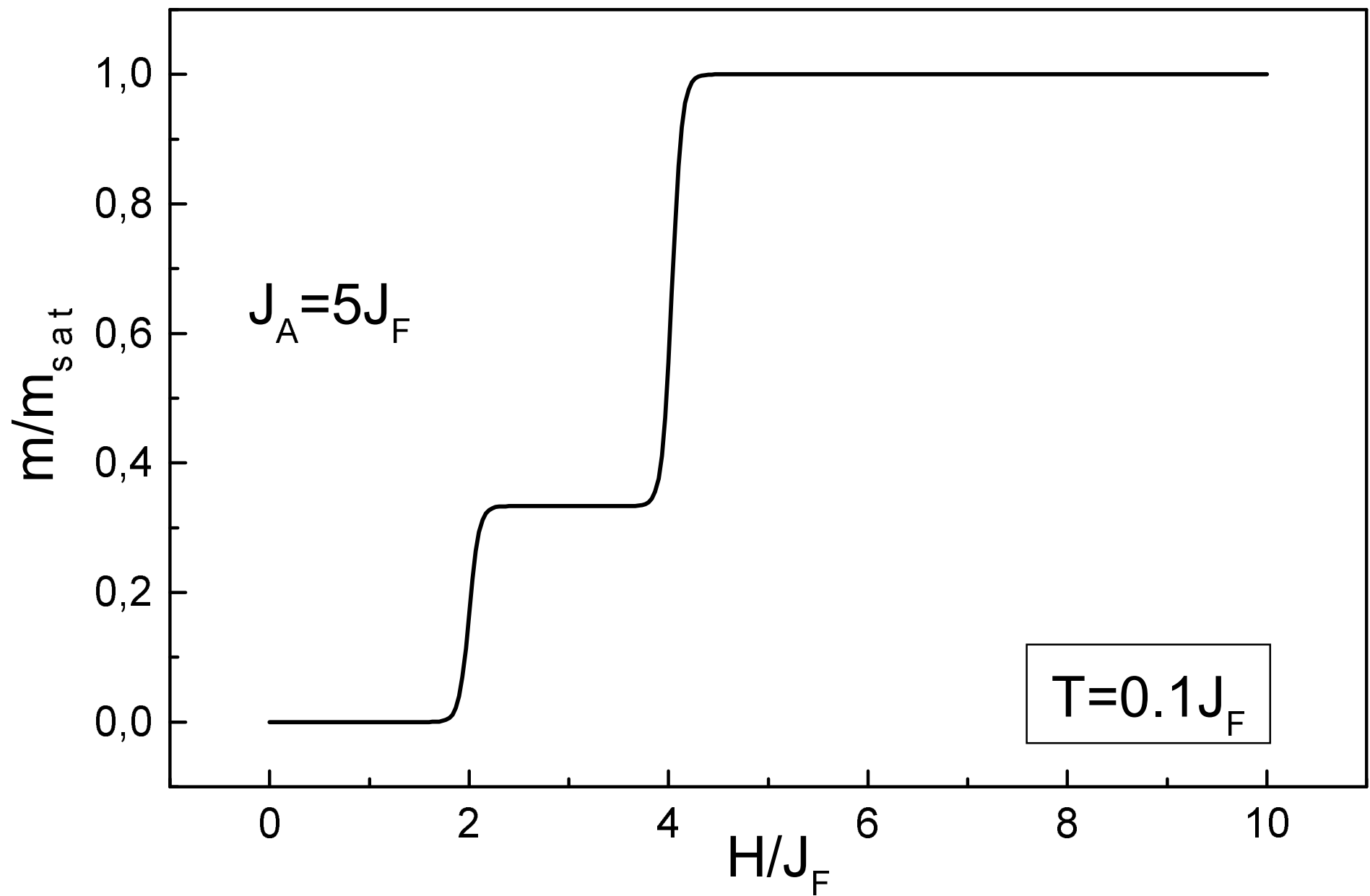
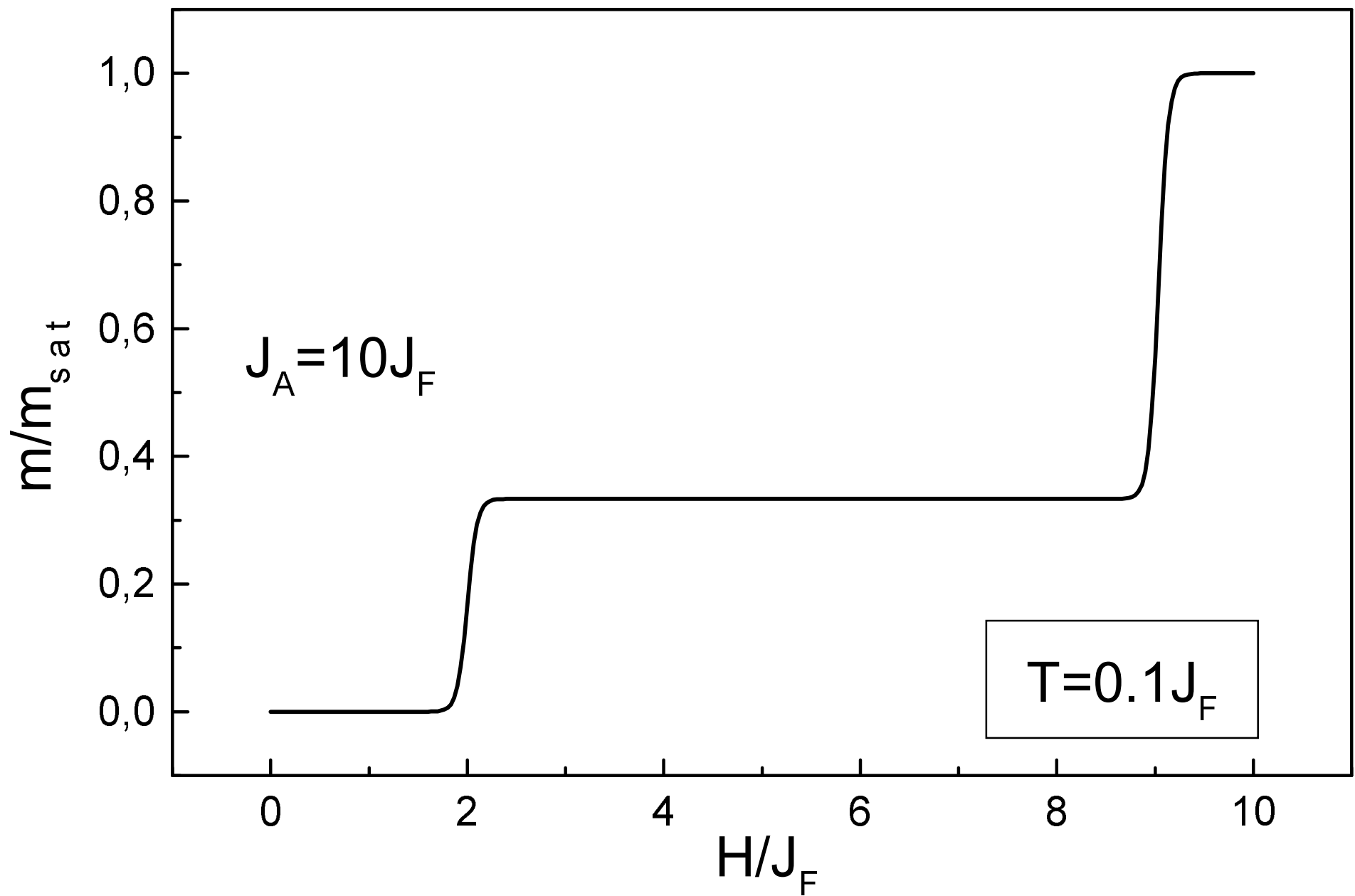
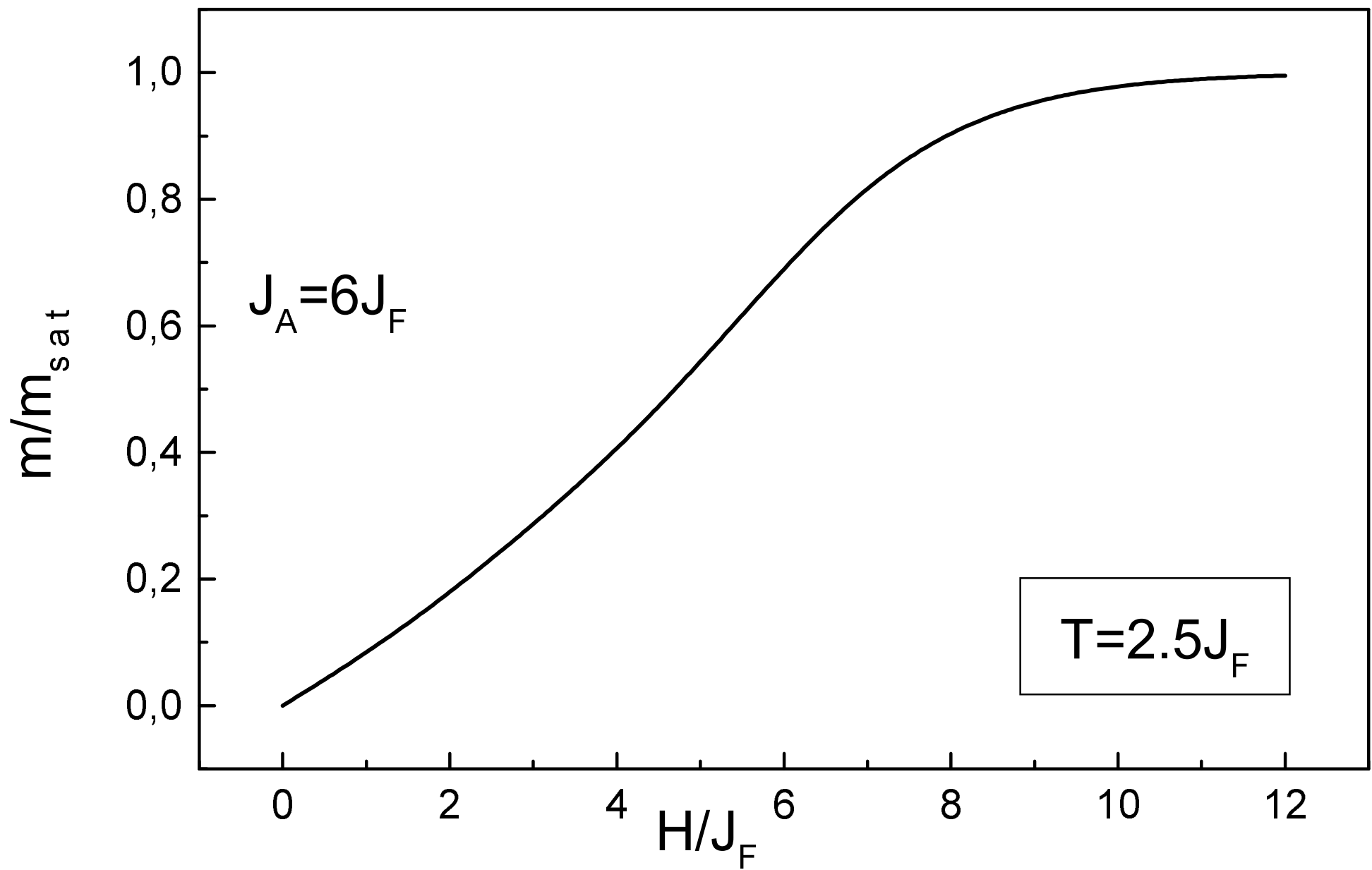
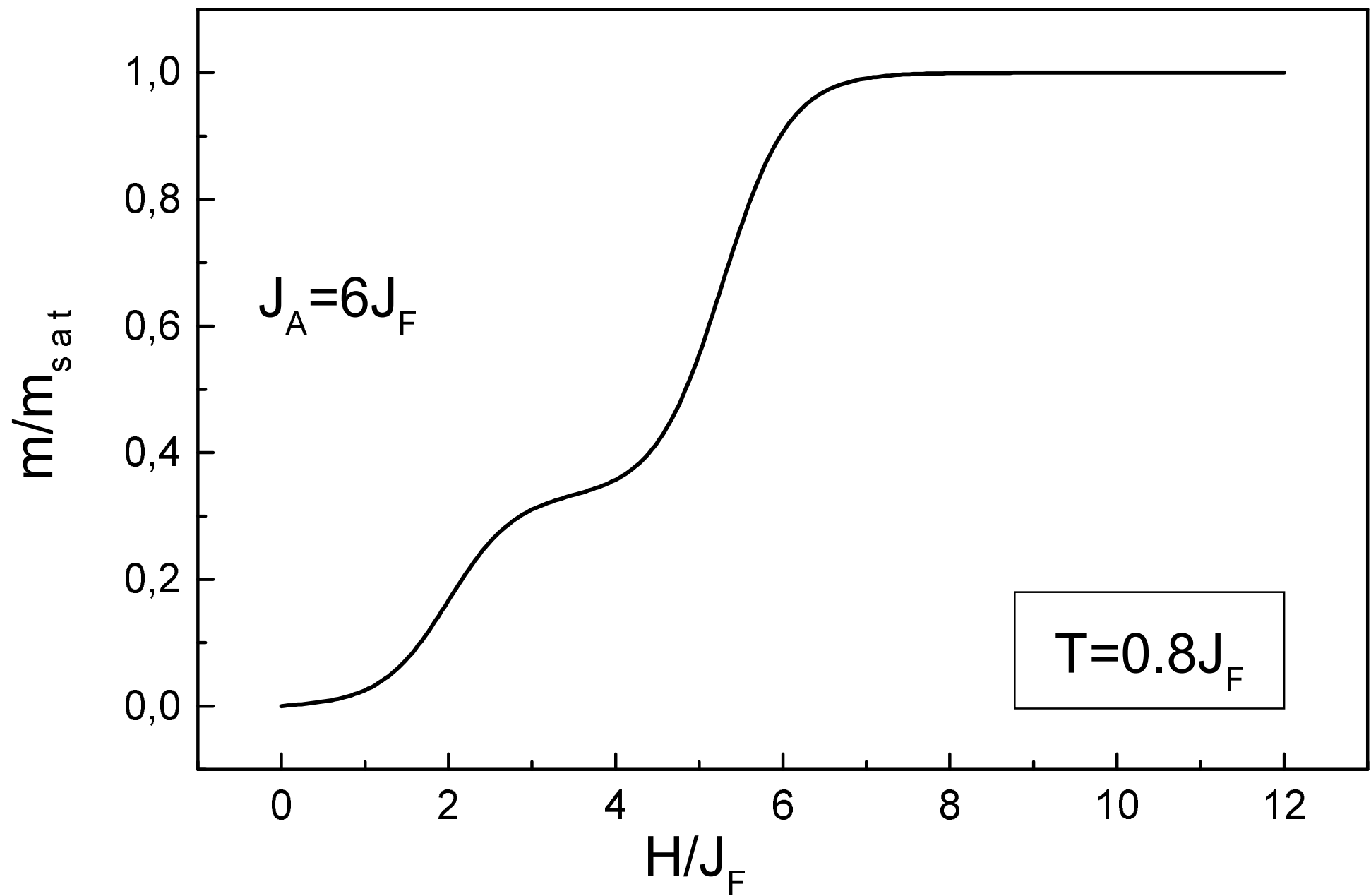
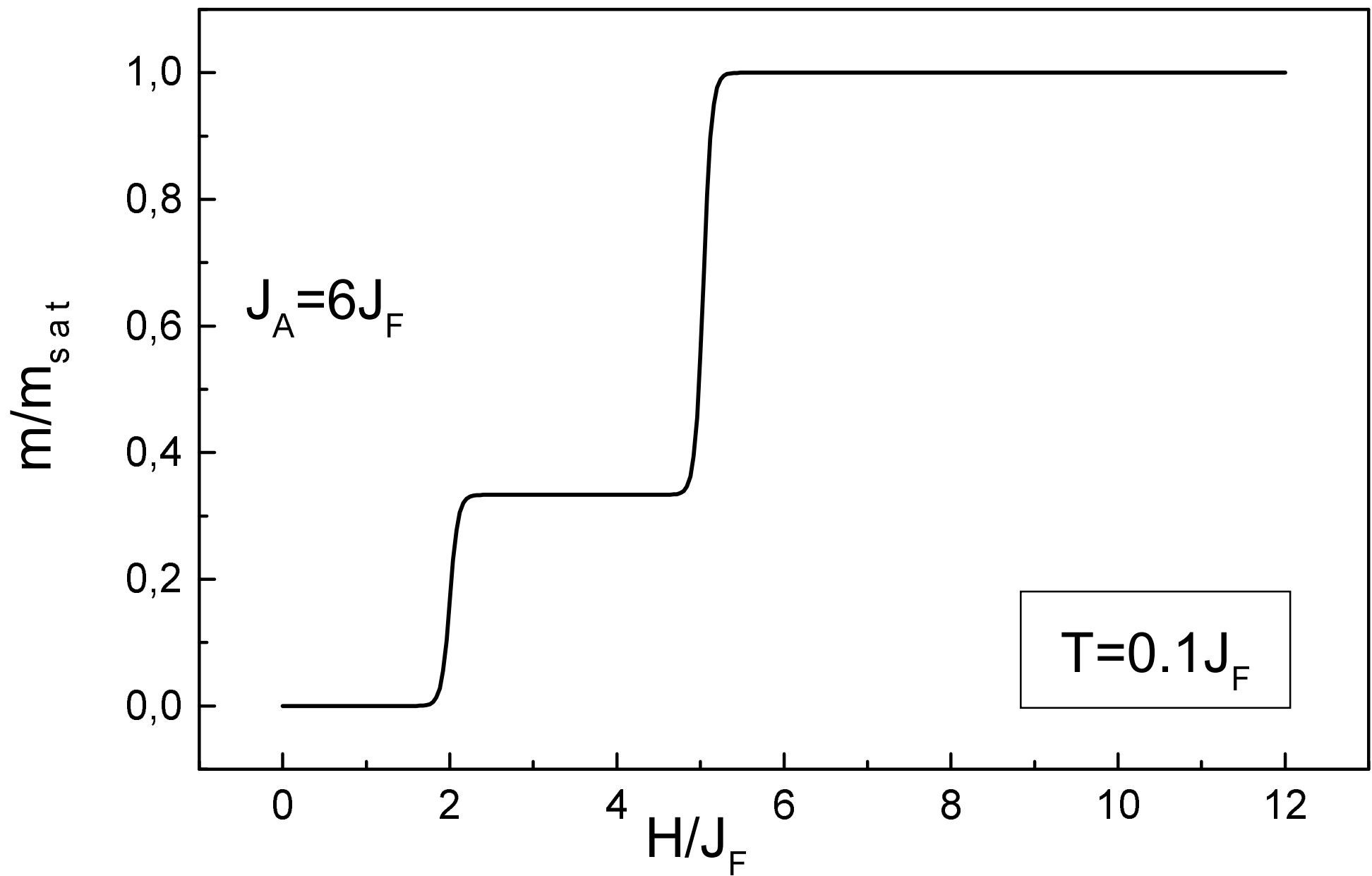
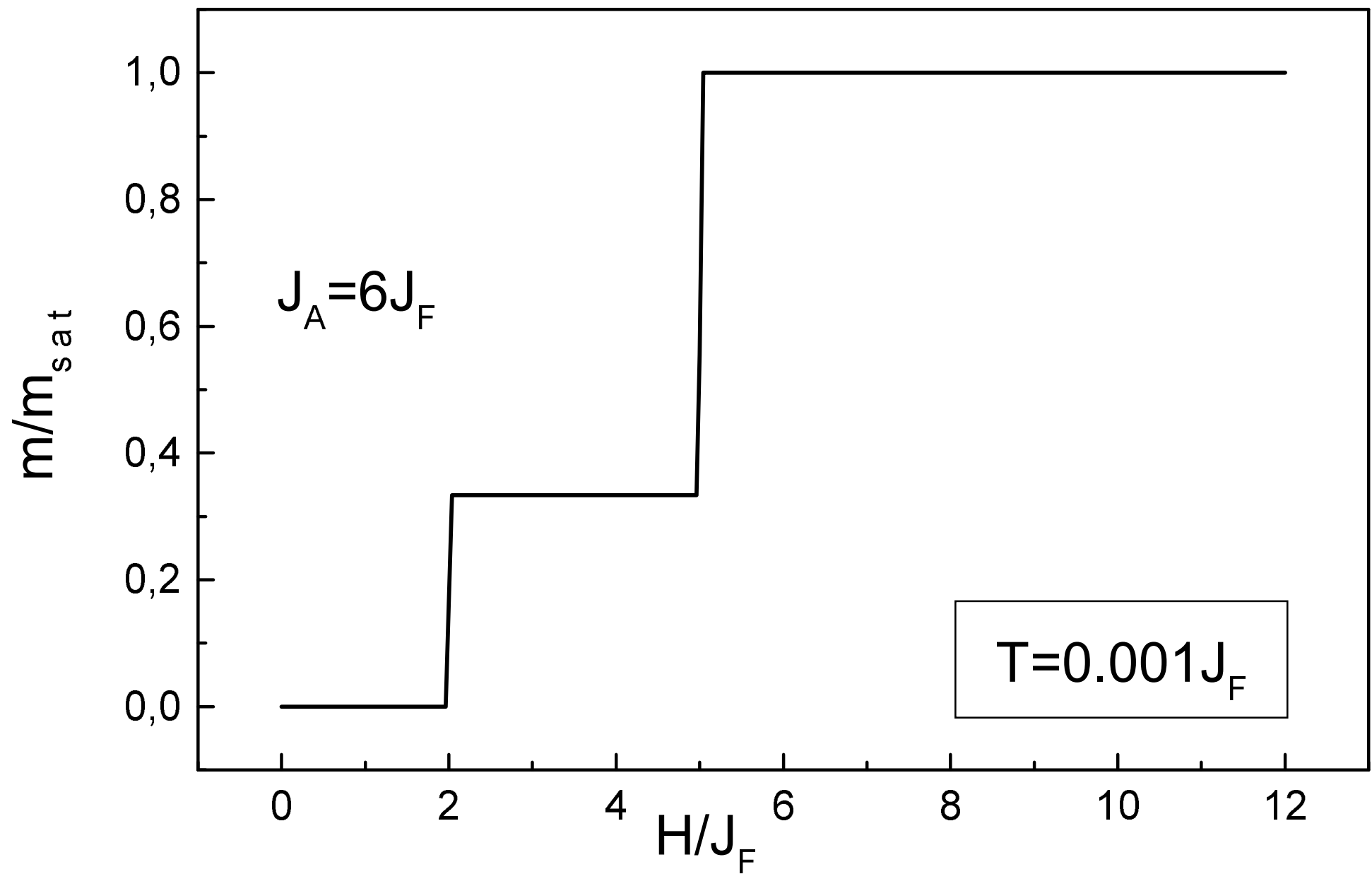
\end{document}